\documentclass[conference]{IEEEtran}
\IEEEoverridecommandlockouts

\usepackage{cite}
\usepackage{amsmath,amssymb,amsfonts,mathtools}
\usepackage{amsthm}
\usepackage{graphicx}
\usepackage{textcomp}
\usepackage{xcolor}
\usepackage{booktabs,array,multirow}
\usepackage{enumitem}
\usepackage{algorithm}
\usepackage{algpseudocode}
\usepackage{tikz}
\usepackage[hidelinks]{hyperref}

\usetikzlibrary{arrows.meta,positioning,shapes,fit,calc}

\newtheorem{theorem}{Theorem}

\newtheorem{proposition}{Proposition}
\newtheorem{corollary}{Corollary}
\newtheorem{assumption}{Assumption}
\newtheorem{definition}{Definition}
\newtheorem{remark}{Remark}

\newcommand{\N}{\mathcal N}
\newcommand{\E}{\mathcal E}

\newcommand{\C}{\mathcal C}

\newcommand{\Z}{\mathcal Z}
\newcommand{\M}{\mathcal M}
\newcommand{\Y}{\mathcal Y}
\newcommand{\A}{\mathcal A}

\newcommand{\U}{\mathcal U}

\newcommand{\Lam}{\Lambda}

\newcommand{\one}{\mathbf 1}
\newcommand{\EE}{\mathbb E}

\newcommand{\ZZ}{\mathbb Z}

\newcommand{\sfm}{\mathrm{sfm}}
\newcommand{\mabp}{\mbox{\normalfont\textsc{MABP}}}

\begin{document}

\title{Augmented Backpressure for Decentralized Management of Agentic Networks
}

\author{
\IEEEauthorblockN{Zuyuan Zhang}
\IEEEauthorblockA{
\textit{The George Washington University}\\
zuyuan.zhang@gwu.edu
}
\and
\IEEEauthorblockN{Sizhe Tang}
\IEEEauthorblockA{
\textit{The George Washington University}\\
s.tang1@gwu.edu
}
\and
\IEEEauthorblockN{Tian Lan}
\IEEEauthorblockA{
\textit{The George Washington University}\\
tlan@gwu.edu
}
}

\maketitle

\begin{abstract}
Agentic foundation-model service networks handle requests spanning retrieval, planning, generation, verification, and tool use. Unlike traditional communication networks, control performance depends on queue dynamics and contextual memory states, including prefix/KV blocks, retrieved contexts, expert warm states, and verified tool outputs. These states arise from execution history and alter service work and downstream successor laws under finite local budgets. Treating them as passive caches or an independent process leaves a queueing-control gap. To this end, we propose \emph{Memory-Augmented Backpressure} (MABP), a queue--memory control framework for stateful foundation-model service networks (SFMSNs) that jointly models commodity queues and causal contextual memory dynamics. MABP represents each request by service and state types, estimates memory-dependent work, penalties, and successor probabilities, then reads queues and resident memory each slot, selects feasible routing, transfer, activation, and service actions using a memory-dependent pressure score, and retains a budget-feasible subset of resident and newly generated objects. We prove an occupation-measure capacity outer bound with conditional tightness. We show that modeling contextual memory can strictly increase the stability region through work reduction and transition shaping, establishing a separation between memory-aware and memory-oblivious decisions. We also prove throughput and drift-plus-penalty guarantees for exact frame-MABP with bounded-loss extensions to approximate solvers. 
\end{abstract}


\section{Introduction}
\label{sec:introduction}

Backpressure and max-weight scheduling stabilize stochastic networks by assigning service and routing decisions to actions with large queue-differential reward~\cite{tassiulas1990stability,neely2010stochastic,georgiadis2006resource,stolyar2004maxweight,dai2005maximum,zhang2026counterfactual,zou2024distributed,yang2026dig}. They are among the standard frameworks for decentralized stochastic network control. In canonical formulations, the controller changes routing, service rates, or resource allocations, while service-generated contextual states are either absent, treated as exogenous cache contents, or not deliberately preserved as part of the stabilizing control.

This abstraction is increasingly inadequate for agentic foundation-model service networks, where a request may traverse retrieval, planning, generation, expert execution, verification, tool use, and rethinking stages across distributed workers~\cite{bommasani2021opportunities,lewis2020retrieval,shazeer2017outrageously,fedus2022switch,yu2022orca,kwon2023efficient,zheng2024sglang,gao2024cost,li2025continuum,yu2026interactive}. These stages leave persistent execution states---retrieved contexts, prefix/KV blocks, expert warm states, activation signatures, and verified tool outputs---that may outlive the request that created them. They are not merely passive cache files. They are produced by prior executions/services, consume local memory, 
and can change how later tasks move through the queueing network. Thus the decentralized control problem for agentic service networks must jointly address two coupled dynamics: the visible queueing dynamics of service demand and the contextual memory-state dynamics generated by executions. Existing queue-only backpressure rule can stabilize the visible queues, while leading to suboptimal contextual memory states and causing unmodeled execution-state storage to grow without bound. Proposition~\ref{prop:state-growth} gives an illustrative example. This motivates modeling contextual memory as a controlled state variable together with queueing dynamics, mandating an augmented backpressure policy.

We introduce a \emph{stateful foundation-model service network} (SFMSN). Each commodity identifies a request class, a service stage, and a finite state type. Each node maintains both commodity queues and a finite contextual memory state. The transitions are causal: a node inherits memory, transfers or activates existing objects, performs routing and service, observes service-generated objects, and retains a budget-feasible subset for the next slot. This differs from conventional multi-resource queueing models 
as active memory control is coupled with queueing dynamics and changes the service operator itself. It can reduce work through \(w_i(a,Y_i)\), alter quality-adjusted penalty through \(p_i(a,Y_i)\), and, most importantly, change the downstream successor law \(P_i(c'\mid a,Y_i)\). Two workers with the same queue lengths and nominal compute capacity can therefore attain different performance and create different downstream loads because they hold different contextual memory states.

The coupled model creates two new challenges for backpressure based control. First, contextual memory states must be captured as routing to a lightly loaded but state-mismatched worker can increase future verifier or tool queues.
Second, memory retention is a delayed control decision. An object generated in slot \(t\) cannot be used until a later request, and an object with little immediate value may be necessary to sustain a favorable recurrent memory state. A one-slot cache bonus is therefore insufficient for a theorem-level stability result.

Memory-Augmented Backpressure (\mabp) prices a service action by the queue it consumes, the downstream queues it is expected to create under the active memory state, and a quality-adjusted penalty. For a service action \(a\) at node \(i\) with active memory \(Y_i=Y_i^{\rm act}(t)\), the local service pressure is
\begin{equation}
\begin{split}
    W_i^a(t;Y_i)
    ={}&Q_i^{\iota(a)}(t)
    -\sum_{c'\in\C}P_i(c'\mid a,Y_i)Q_i^{c'}(t)\\
    &-\nu p_i(a,Y_i).
\end{split}
    \label{eq:intro-pressure}
\end{equation}
The first term models input-backlog consumption; the second prices expected downstream congestion and depends on the active memory state through \(P_i(c'\mid a,Y_i)\); and the third controls the penalty--delay tradeoff and depends on memory through \(p_i(a,Y_i)\).
Memory-dependent work \(w_i(a,Y_i)\) enters the compute constraint, so the scheduler chooses a feasible combination of routing, transfer, activation, and service actions with large total pressure rather than simply adding a cache-hit reward. Next, generated and resident objects are evaluated by a retention rule that uses the pre-service queue vector, preserving causality. 
A slot is one discrete scheduling interval in which multiple queued requests may be routed or served. The exact policy used in the analysis is an \(H\)-slot frame oracle that plans over controlled memory states.
The one-slot rule used in experiments is its deployable surrogate. This paper goes beyond research on LLM systems that often focus on batching, prefix/KV reuse, phase splitting, cache affinity, and load-aware request placement~\cite{li2023alpaserve,kwon2023efficient,agrawal2024taming,zhong2024distserve,patel2024splitwise,gao2024cost,li2025continuum}, rather than the stochastic, multi-stage stability region shaped jointly by queues and contextual memory states.

We rigorously analyze the proposed scheme. Theorem~\ref{thm:capacity} gives an occupation-measure outer bound on the stability region and proves conditional tightness under a queue-compatible realization condition, making explicit that recurrent memory states must be generated and sustained causally. We prove that contextual memory can strictly increase the capacity region through two independent mechanisms: reducing service work and reducing downstream verifier/tool creation (in Proposition~\ref{prop:work-containment}, Theorem~\ref{thm:transition-containment}, and Corollary~\ref{cor:prob-transition}). We also characterize a minimax one-step gap for decisions that observe queues but ignore memory states (in Proposition~\ref{prop:separation}) and quantify throughput for exact frame-\mabp (in Theorem~\ref{thm:throughput}) and a drift-plus-penalty tradeoff (in Theorem~\ref{thm:dpp}). An approximate frame solver using a surrogate is shown to inherit the same guarantees (in Corollary~\ref{cor:additive-approx}).

The paper makes three contributions.
\begin{itemize}[leftmargin=1.2em,itemsep=0.2em]
    \item \textbf{Coupled queue--memory model.} We introduce SFMSN. 
    It couples queue dynamics with memory-state dynamics as active memory in agentic networks can alter service work, quality-adjusted penalty, and downstream execution.
    \item \textbf{Stateful capacity and separation theory.} We derive occupation-measure constraints that are necessary for stability and conditionally tight under an explicit queue-compatible realization condition. We also prove strict capacity enlargement from work reuse and from memory-dependent transition laws, and we give a minimax separation for memory-oblivious decisions.
    \item \textbf{Memory-Augmented Backpressure.} We formulate exact frame-\mabp, which plans over causal memory states and achieves throughput and drift-plus-penalty guarantees for strictly feasible arrivals. We then give a practical one-slot surrogate based on the same memory-dependent pressure and specify the bounded-loss condition under which an approximate solver preserves the frame guarantees.
\end{itemize}

\begin{figure}[t]
\centering
\begin{tikzpicture}[>=Latex,node distance=5mm and 6mm,
                    scale=0.72,every node/.style={transform shape}]
\tikzset{
srv/.style={draw,rounded corners,minimum width=1.05cm,
            minimum height=0.58cm,align=center,fill=gray!8},
mem/.style={draw,dashed,rounded corners,minimum width=1.25cm,
            minimum height=0.45cm,align=center,fill=blue!5}
}
\node[srv] (in) {Ingress};
\node[srv,right=of in] (ret) {Retrieve};
\node[srv,right=of ret] (gen) {Generate};
\node[srv,right=10mm of gen] (out) {Egress};
\node[srv,below=8mm of out] (ver) {Verify/Tool};
\node[mem,above=5mm of gen] (kv) {prefix/KV\\lowers \(w_i\)};
\node[mem,below=10mm of ret] (cert) {verified result\\reduces \(P_i(c_{\rm v}\mid a,Y_i)\)};

\draw[->] (in) -- (ret);
\draw[->] (ret) -- (gen);
\draw[->] (gen) -- node[above]{direct} (out);
\draw[->] (gen) |- node[pos=0.25,left]
    {\scriptsize \(P_i(c_{\rm v}\mid a,Y_i)\)} (ver);
\draw[->] (ver) -- (out);
\draw[->,dashed] (kv) -- (gen);
\draw[->,dashed] (cert) -| (gen);
\end{tikzpicture}
\caption{A stateful service trajectory. Different execution states affect different primitives: prefix/KV state reduces current work, while a reusable verified result can reduce the probability of creating a downstream verifier or tool job.}
\label{fig:running-example}
\vspace{-2mm}
\end{figure}
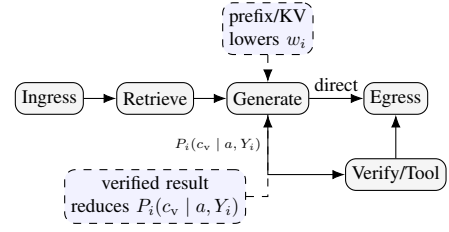

\section{Model}
\label{sec:model}

This section defines a causal stateful-service model. In each slot, inherited memory is transferred and activated before routing and service; completed service creates successor jobs and memory only for later slots. Thus memory affects work, transition laws, and quality-adjusted penalties without same-slot lookahead. Catalog caching is recovered when objects alter only access cost, while result caching is service-generated execution memory~\cite{kamran2021deco,mao2017survey,ndikumana2019joint,maia2024survey}.

\subsection{Network and Stateful Commodities}

The service system is a finite directed graph $\mathcal G=(\N,\E)$. Link $(i,j)$ has capacity $C_{ij}$ and per-unit penalty $d_{ij}^c$ for commodity $c$; node $i$ has compute capacity $\mu_i$ and memory budget $B_i$. Time is slotted, and routing/service amounts are bounded integer quanta, giving a finite approximation to continuous rates.

A commodity is $c=(r,k,z)\in\C$, where $r$ is the request class, $k\in\{0,\ldots,L_r-1\}$ is the service stage, and $z\in\Z_{r,k}$ is a finite state type encoding, e.g., a prompt cluster, retrieval signature, expert-subset signature, prefix/KV signature, or tool-result signature~\cite{lewis2020retrieval,shazeer2017outrageously,fedus2022switch,kwon2023efficient,zheng2024sglang,li2025continuum}. Queue $Q_i^c(t)$ stores unserved units of $c$ at node $i$. Let $\mathcal I\subseteq\N\times\C$ be the finite set of admissible node--commodity queues, retained only when a feasible routing or service path can drain them. We set $Q_i^c(t)=0$ for $(i,c)\notin\mathcal I$, and all sums over $i,c$ range over $\mathcal I$ unless stated otherwise.

Let $\mathcal H_t$ be the pre-decision history containing $Q(t)$, inherited memory $Y^-(t)$, and past controls/outcomes, but not slot-$t$ arrivals or service-generated outcomes. All slot-$t$ controls are $\mathcal H_t$-measurable. Exogenous arrivals join $Q(t+1)$, satisfy $A_i^c(t)\in\ZZ_+$, $\EE[A_i^c(t)\mid\mathcal H_t]=\lambda_i^c$, and a uniform conditional second-moment bound; write $\lambda=(\lambda_i^c)_{(i,c)\in\mathcal I}$. They are conditionally independent of controlled service and memory-generation randomness given $\mathcal H_t$.

\subsection{A Running Example}

Fig.~\ref{fig:running-example} illustrates the service trajectory. A request is matched with retrieved context, routed to a generator that may hold a compatible prefix/KV or expert state, and then completed or verified. Each completed stage may generate an object for later requests. Classical backpressure observes queues and nominal capacities; \mabp\ additionally prices how the resident execution state changes service work and downstream job creation.

\subsection{Causal Execution-State Memory}
\label{sec:memory-dynamics}

Let $\M$ be the finite set of signature--replica identities. Each object $m\in\M$ has size $b_m>0$ and represents a bounded content signature with a replica/version slot or a refreshed local instance. Feasible global memory states are
$
\Y=\{(Y_i)_{i\in\N}:\,Y_i\subseteq\M,\ 
\sum_{m\in Y_i}b_m\le B_i\ \forall i,\ 
Y_i\cap Y_j=\emptyset\ \forall i\ne j\}.
$
Objects with the same content signature may share primitive parameters, but simultaneous replicas have distinct identities. Object $m$ has compatibility indicator $\chi_m(c,a)\in\{0,1\}$; for action $a$, $Y_i[a]=\{m\in Y_i:\chi_m(\iota(a),a)=1\}$ is the compatible active subset. The primitives $w_i(a,Y_i)$, $p_i(a,Y_i)$, and $P_i(\cdot\mid a,Y_i)$ depend on memory only through this subset, which is implemented as a finite compatibility signature indexing profiled primitive tables.

At the beginning of slot $t$, node $i$ inherits $Y_i^-(t)$. A transfer decision uses indicators $u_{ij}^m(t)\in\{0,1\}$ with
$
u_{ij}^m(t)\le \one\{m\in Y_i^-(t)\},\qquad
\sum_{j:(i,j)\in\E}u_{ij}^m(t)\le 1 .
$
With $M_i^{\rm out}(t)=\{m:u_{ij}^m(t)=1\text{ for some }j\}$ and $M_i^{\rm in}(t)=\{m:u_{ji}^m(t)=1\text{ for some }j\}$, the resident set is $R_i(t)=(Y_i^-(t)\setminus M_i^{\rm out}(t))\cup M_i^{\rm in}(t)$. The transfer vector $u(t)\in\U(Y^-(t))$ enforces transfer bandwidth, topology, destination-budget, global-identity, and any joint data/state link-capacity constraints. The base model is migration; replication can be represented by assigning a copied object a globally free replica identity, without affecting the results below.

Resident and active memory satisfy
\begin{equation}
\sum_{m\in R_i(t)}b_m\le B_i,\qquad
Y_i^{\rm act}(t)\subseteq R_i(t).
\label{eq:memory-budget}
\end{equation}
Activation-bandwidth, device-residency, and concurrent-state limits are included in the feasible control set. Slot-$t$ generated memory cannot be used in that slot. After service, completed actions generate random sets $G_i(t)\subseteq\M$ respecting global identity constraints; a generated identity must be globally free unless it refreshes an identity already resident at node $i$.

Before observing $G(t)$, the controller selects causal retention maps $E(t)=(E_i(t))_{i\in\N}$, where $E_i(t):2^{\M}\to2^{\M}$ and $E_i(t)(A)\subseteq A$. For every generated set with positive probability, the retained vector must satisfy the budgets and belong to $\Y$:
\begin{equation}
Y_i^+(t)=E_i(t)\bigl(R_i(t)\cup G_i(t)\bigr),
\quad
Y_i^-(t+1)=Y_i^+(t).
\label{eq:memory-dynamics}
\end{equation}
Using $R_i(t)$, rather than $Y_i^{\rm act}(t)$, keeps inactive resident objects eligible for retention instead of forcing their same-slot eviction.

Let $\gamma(t)=\bigl(u(t),Y^{\rm act}(t),F(t),S(t),E(t)\bigr)\in\Gamma(Q(t),Y^-(t))$ denote the complete feasible one-slot control, including the routing and service variables below. All components are chosen before service outcomes and generated objects are observed; after $G(t)$ is realized, only the preselected maps $E_i(t)(\cdot)$ are evaluated, so retention is adaptive to generated objects but nonanticipative.

A representative memory cost is
$
h^{\rm mem}(t)=
\sum_{(i,j)\in\E}\sum_{m\in\M}\kappa_{ij}^m u_{ij}^m(t)
+\sum_i\theta_i\sum_{m\in Y_i^+(t)}b_m
+\sum_i\sum_{m\in G_i(t)\cap Y_i^+(t)}\omega_i^m ,
$
where $\kappa_{ij}^m,\theta_i,\omega_i^m\ge0$ are transfer, holding, and write/update costs. Define $\bar h^{\rm mem}(y,\gamma)=\EE[h^{\rm mem}(t)\mid Y^-(t)=y,\gamma(t)=\gamma]$. Conditioned on $Y^-(t)=y$ and feasible $\gamma$, the laws of successor quanta and generated memory are determined by the state--control pair and may be correlated across simultaneous completions. Since $\M$ and $\Y$ are finite, generation and causal retention induce the controlled memory kernel $\mathsf T(y'\mid y,\gamma)=\Pr\{Y^-(t+1)=y'\mid Y^-(t)=y,\gamma(t)=\gamma\}$, which depends on $Q(t)$ only through the selected queue-feasible control $\gamma$.

\subsection{Routing and Service Actions}

The controller chooses routing amounts $F_{ij}^c(t)\in\ZZ_+$ satisfying $\sum_{c\in\C}F_{ij}^c(t)\le C_{ij}$ for each $(i,j)\in\E$. Routing preserves commodity labels, and $F_{ij}^c(t)=0$ unless both $(i,c)$ and $(j,c)$ belong to $\mathcal I$; incoming routed traffic becomes available in the next slot.

Node $i$ has a finite action set $\A_i$. Each action $a\in\A_i$ consumes commodity $\iota(a)$, and actions with inadmissible input queues are omitted. Under active memory $Y_i^{\rm act}(t)$, action $a$ requires work $w_i(a,Y_i^{\rm act}(t))>0$ per completed unit, with uniformly bounded work values by finiteness. Completed service amounts $S_i^a(t)\in\ZZ_+$ satisfy $\sum_{a\in\A_i}S_i^a(t)w_i(a,Y_i^{\rm act}(t))\le \mu_i$. Routing and service are jointly queue-feasible:

\begin{equation}
\sum_{j:(i,j)\in\E}F_{ij}^c(t)
+\sum_{a\in\A_i:\iota(a)=c}S_i^a(t)
\le Q_i^c(t),\qquad (i,c)\in\mathcal I .
\label{eq:queue-feasibility}
\end{equation}

A completed unit of action $a$ either exits or creates one downstream commodity $c'$ with probability $P_i(c'\mid a,Y_i^{\rm act}(t))$. With absorbing departure state $\varnothing$ and $Q_i^\varnothing=0$, $\sum_{c'\in\C\cup\{\varnothing\}}P_i(c'\mid a,Y_i)=1$; the support is restricted to $\varnothing$ and commodities $c'$ with $(i,c')\in\mathcal I$. A successor job joins $Q_i^{c'}(t+1)$ and can be routed or served only later.

The quality-adjusted service penalty is $p_i(a,Y_i)=\ell_i(a,Y_i)-\eta v_i(a,Y_i)$, $\eta\ge0$, where $\ell_i$ is an operator-defined cost or quality-loss term and $v_i$ is a reuse/quality credit not already captured by $w_i$. A service-time term in $\ell_i$ must not duplicate work represented by $w_i$. Penalties are uniformly bounded and may be negative; $\eta$ weights the credit, while $\nu$ controls the queue--penalty tradeoff.

The total one-slot quality-adjusted penalty is
$
h(t)=
\sum_{(i,j)\in\E}\sum_{c\in\C}d_{ij}^cF_{ij}^c(t)
+\sum_i\sum_{a\in\A_i}p_i(a,Y_i^{\rm act}(t))S_i^a(t)
+h^{\rm mem}(t).
$
Its conditional expectation is obtained by replacing $h^{\rm mem}(t)$ with $\bar h^{\rm mem}(y,\gamma)$ and is written as $J(y,\gamma)$ in Section~\ref{sec:capacity}. Since feasible routing, service, and memory controls are bounded and primitive penalties are uniformly bounded, $h(t)$ is uniformly bounded above and below over one slot.

\subsection{Queue Dynamics and Stability}

For each $(i,c)\in\mathcal I$,
\begin{align}
Q_i^c(t+1)
&=Q_i^c(t)-\sum_{j:(i,j)\in\E}F_{ij}^c(t)
-\sum_{a:\iota(a)=c}S_i^a(t)
+A_i^c(t) \notag\\
&\quad+\sum_{j:(j,i)\in\E}F_{ji}^c(t)
+\sum_{a\in\A_i}\xi_i^{a,c}(t),
\label{eq:queue-dynamics}
\end{align}
where $\EE[\xi_i^{a,c}(t)\mid Q(t),Y^-(t),\gamma(t)]
=S_i^a(t)P_i(c\mid a,Y_i^{\rm act}(t))$. Nonnegativity follows from~\eqref{eq:queue-feasibility}. Also, $\xi_i^{a,c}(t)\in\ZZ_+$ and $\sum_{c:(i,c)\in\mathcal I}\xi_i^{a,c}(t)\le S_i^a(t)$ almost surely, so each completion creates at most one successor job. Bounded routing/service quanta and the arrival second-moment assumption then give a uniform conditional second-moment bound on one-slot queue increments.

A policy is admissible if it is causal with respect to $\{\mathcal H_t\}$ and satisfies all routing, compute, transfer, memory, and queue-feasibility constraints. The network is strongly stable if $\limsup_{T\to\infty}T^{-1}\sum_{t=0}^{T-1}\sum_{(i,c)\in\mathcal I}\EE[Q_i^c(t)]<\infty$. For initial memory state $y^0\in\Y$, the stateful capacity region $\Lam_{\sfm}(y^0)$ is the closure of all $\lambda\in\mathbb R_+^{|\mathcal I|}$ for which some admissible policy strongly stabilizes all queues from every finite initial queue vector and $Y^-(0)=y^0$; write $\Lam_{\sfm}$ when $y^0$ is fixed or irrelevant. The no-execution-memory region $\Lam_{\rm no\text{-}mem}$ forces $Y_i^-(t)=Y_i^{\rm act}(t)=Y_i^+(t)=\emptyset$ for all $i,t$, evaluates all primitives at empty memory, and discards generated objects. A policy is memory-oblivious if routing and service depend on queue history and fixed system parameters, but not on realized $Y^-(t)$, and its service scores are invariant to replacing the realized execution-memory state by any other feasible memory state.

\section{Stateful Capacity and Separations}
\label{sec:capacity}

This section characterizes capacity under controlled execution memory, which must be causally generated, retained, and revisited. We define an occupation-measure outer region, state its realization condition, and isolate capacity gains from work reuse and successor-law shaping.

Fix \(y^0\). For inherited memory \(y\), let \(\Gamma(y)\) be the finite set of one-slot controls satisfying routing, compute, transfer, activation, and memory constraints before queue truncation, and let \(\Gamma(q,y)\subseteq\Gamma(y)\) be the queue-feasible subset. A control \(\gamma\) specifies transfer, active memory, routing, service, and retention; \(Y_i^\gamma\) is the active memory at node \(i\), and \(\mathsf T(y'\mid y,\gamma)\) is the induced memory kernel. Let \(\Y_{\rm r}(y^0)\) be the states reachable from \(y^0\). Occupation measures use \(\Gamma(y)\); queue-compatible achievability is imposed by Assumption~\ref{ass:realization}.

A stationary memory-state occupation measure is a nonnegative collection \(x(y,\gamma)\), with \(x(y,\gamma)=0\) for \(y\notin\Y_{\rm r}(y^0)\), satisfying
\begin{align}
&\sum_{y\in\Y}\sum_{\gamma\in\Gamma(y)}x(y,\gamma)=1,
\label{eq:occ-normalization}\\
&\sum_{\gamma\in\Gamma(y)}x(y,\gamma)
 =\sum_{\tilde y\in\Y}\sum_{\tilde\gamma\in\Gamma(\tilde y)}
 x(\tilde y,\tilde\gamma)\mathsf T(y\mid\tilde y,\tilde\gamma),
 \quad y\in\Y .
\label{eq:memory-stationarity}
\end{align}
The second constraint is the memory-flow balance; it prevents recurrent memory configurations from being treated as static cache placements.

For a state--control pair, define the expected net service surplus of queue \((i,c)\) by
\begin{align}
g_i^c(y,\gamma)
={}&\sum_{j:(i,j)\in\E}F_{ij}^c(\gamma)
 +\sum_{a:\iota(a)=c}S_i^a(\gamma)
-\sum_{j:(j,i)\in\E}F_{ji}^c(\gamma) \notag\\
&-\sum_{a\in\A_i}S_i^a(\gamma)P_i(c\mid a,Y_i^\gamma).
\label{eq:net-service}
\end{align}
Whenever \(\gamma(t)\) is queue-feasible,
\(\EE[Q_i^c(t+1)-Q_i^c(t)\mid Q(t),Y^-(t),\gamma(t)]
=\lambda_i^c-g_i^c(Y^-(t),\gamma(t))\).

The conditional expected one-slot penalty is \(J(y,\gamma)=
\sum_{(i,j)\in\E}\sum_{c\in\C}F_{ij}^c(\gamma)d_{ij}^c
+\sum_{i\in\N}\sum_{a\in\A_i}S_i^a(\gamma)p_i(a,Y_i^\gamma)
+\bar h^{\rm mem}(y,\gamma)\). Let
\(\bar J(x)=\sum_{y,\gamma}x(y,\gamma)J(y,\gamma)\), and assume per-slot penalties are uniformly bounded by \(J_{\min}\) and \(J_{\max}\).

An occupation measure supports \(\lambda\) with slack \(\delta\ge0\) if
\begin{equation}
\bar g_i^c(x)\triangleq
\sum_{y\in\Y}\sum_{\gamma\in\Gamma(y)}x(y,\gamma)g_i^c(y,\gamma)
\ge \lambda_i^c+\delta,
\quad (i,c)\in\mathcal I .
\label{eq:capacity-balance}
\end{equation}

\begin{definition}[Occupation-measure outer region]
\label{def:occupation-region}
\(\Lam_{\rm occ}(y^0)\) is the closure of all arrival vectors satisfying
\eqref{eq:occ-normalization}--\eqref{eq:capacity-balance}
for some occupation measure supported on \(\Y_{\rm r}(y^0)\). It is strictly feasible if the inequalities in~\eqref{eq:capacity-balance} hold with a uniform \(\delta>0\).
\end{definition}

The occupation region need not be achievable: an empty-queue truncation can change both current service surplus and the future memory state. The next assumption certifies a queue-compatible realization of the occupation measure.

\begin{assumption}[Queue-compatible realization with bounded transient loss]
\label{ass:realization}
For every reachable occupation measure \(x\) supporting \(\lambda\) with positive slack, there exist \(H_x\in\ZZ_+\) and \(L_x\ge0\) such that, for every \(H\ge H_x\), there are finite \(C_{H,x},D_{H,x}\ge0\) and a causal \(H\)-slot rule with \(\gamma(\tau)\in\Gamma(Q(\tau),Y^-(\tau))\). Writing \(q\cdot v=\sum_{(i,c)\in\mathcal I}q_i^c v_i^c\), uniformly over frozen queues \(q\) and reachable initial memory states \(y\), initialization with \(Q(0)=q\) and \(Y^-(0)=y\) satisfies
\begin{align}
\EE\!\left[
\sum_{\tau=0}^{H-1}q\cdot g(Y^-(\tau),\gamma(\tau))
\right]
&\ge H\,q\cdot\bar g(x)-L_x\|q\|_1-C_{H,x},
\label{eq:uniform-realization}\\
\EE\!\left[
\sum_{\tau=0}^{H-1}J(Y^-(\tau),\gamma(\tau))
\right]
&\le H\bar J(x)+D_{H,x}.
\label{eq:penalty-realization}
\end{align}
\end{assumption}

Assumption~\ref{ass:realization} is stronger than finiteness or unichainness: it excludes memory frequencies that collapse under queue truncation. The frame proof therefore requires \(H\delta>L_x\).

\begin{theorem}[Occupation outer bound and conditional tightness]
\label{thm:capacity}
Under finite memory and control sets and bounded conditional second moments, every strongly stabilizable arrival vector from initial memory \(y^0\) belongs to \(\Lam_{\rm occ}(y^0)\). If strictly feasible vectors are dense in \(\Lam_{\rm occ}(y^0)\) and every strictly feasible vector is supported by a reachable occupation measure satisfying Assumption~\ref{ass:realization}, then
$
    \Lam_{\sfm}(y^0)=\Lam_{\rm occ}(y^0).
$
\end{theorem}
\begin{proof}[Proof sketch]
For necessity, form
\(x_T(y,\gamma)=T^{-1}\sum_{t<T}\Pr\{Y^-(t)=y,\gamma(t)=\gamma\}\).
Compactness gives a convergent subsequence, and strong stability gives a further subsequence with \(\EE[\|Q(T_n)\|_1]/T_n\to0\). Telescoping memory indicators yields~\eqref{eq:memory-stationarity}; telescoping~\eqref{eq:queue-dynamics} yields
\(T_n^{-1}\sum_{t<T_n}\EE[g_i^c(Y^-(t),\gamma(t))]\to\lambda_i^c\), hence~\eqref{eq:capacity-balance} with zero slack and support only on reachable memory states.

For sufficiency, fix a strictly feasible \(x\) with slack \(\delta>0\). Assumption~\ref{ass:realization} supplies a queue-compatible \(H\)-slot comparator. Choosing \(H\) with \(H\delta>L_x\), Theorem~\ref{thm:throughput} gives negative linear frame drift outside a bounded set while~\eqref{eq:penalty-realization} controls penalties. Thus all strictly feasible vectors covered by the assumption are stabilizable; density and closure give equality.
\end{proof}

The next results isolate two independent capacity gains from causally retained execution memory: reduced work and changed successor generation.

\begin{proposition}[Work-reuse strict containment] 
\label{prop:work-containment} 
For the empty initial memory state \(y^\emptyset\), $ \Lam_{\rm no\text{-}mem}\subseteq\Lam_{\sfm}(y^\emptyset), $ and the containment can be strict in a single-stage network. 
\end{proposition}
\begin{proof}[Proof sketch]
Containment follows by discarding all generated objects and emulating the no-memory system. For strictness, take one server with compute capacity \(\kappa_w\ge2\). Without memory, each request consumes \(\kappa_w\) units, so the service rate is \(1\); after one completed request generates and retains a compatible object \(m\), each later request consumes one unit, so the service rate is \(\kappa_w\). The setup phase is finite and does not affect strong stability, hence every \(\lambda\in(1,\kappa_w)\) separates the regions.
\end{proof}

The second gain does not rely on reducing service work. The next example keeps all work requirements fixed and separates the regions only through memory-dependent successor generation.

\begin{theorem}[Transition-law strict containment] 
\label{thm:transition-containment} 
There exists an SFMSN with empty initial memory in which \(w_i(a,Y_i)\) is identical for all memory states but $ \Lam_{\rm no\text{-}mem}\subsetneq\Lam_{\sfm}(y^\emptyset) $ because the successor law \(P_i(c'\mid a,Y_i)\) is memory-dependent. 
\end{theorem}
\begin{proof}[Proof sketch]
Let the upstream server have capacity \(2\), the verifier capacity \(1\), and the link capacity at least \(2\). Work is one unit at each node for all memory states. Without memory, every upstream completion creates a verifier job, so stability requires \(\lambda<1\). With a retained object \(m\), generated after the first upstream completion, later upstream completions depart directly; the verifier sees only finite transient load while the upstream server stabilizes every \(\lambda<2\). Thus any \(\lambda\in(1,2)\) proves strict containment without changing \(w_i(a,Y_i)\).
\end{proof}

\begin{corollary}[Probabilistic verifier reduction]
\label{cor:prob-transition}
Let the upstream and verifier completion capacities be \(\mu_u\) and \(\mu_v\), and assume the upstream--verifier link has capacity at least \(\mu_u\). If the first upstream completion deterministically generates and retains \(m\), and
\(P_u(c_v\mid a,\emptyset)=\pi_h\), \(P_u(c_v\mid a,\{m\})=\pi_l\), with \(0\le \pi_l<\pi_h\le1\), then the no-memory system requires
\(\lambda<\min\{\mu_u,\mu_v/\pi_h\}\), whereas retaining \(m\) permits
\(\lambda<\min\{\mu_u,\mu_v/\pi_l\}\), with \(\mu_v/0=\infty\).
\end{corollary}
\begin{proof}[Proof sketch]
The upstream constraint is \(\lambda<\mu_u\). The verifier receives rate \(\pi_h\lambda\) without memory and \(\pi_l\lambda\) after the finite memory-generation transient. Combining these with the verifier capacity gives the two intervals. The improvement is strict exactly when the reduced verifier-arrival rate relaxes the active bottleneck.
\end{proof}

\section{Memory-Augmented Backpressure}
\label{sec:mabp}

This section derives the queue controller from Section~\ref{sec:capacity}. The key score is the memory-augmented backpressure weight \(\Phi\), whose net-service form matches \(g_i^c\). Since retention may affect only future service, the theorem-level policy is an exact finite-frame oracle; the experiments use a one-slot surrogate that inherits guarantees only under Corollary~\ref{cor:additive-approx}.

A queue-only rule can stabilize visible demand while leaving execution memory uncontrolled:

\begin{proposition}[Queue-stable but unmanaged-memory growth]
\label{prop:state-growth}
A queue-only rule can strongly stabilize a single-server request queue while the expected number of unmanaged service-generated objects grows linearly.
\end{proposition}
\begin{proof}[Proof sketch]
With Bernoulli arrivals \(0<\lambda<1\), unit service, and independent object generation probability \(\rho>0\) per completion, rate conservation gives \(\EE[M(T)]/T\to\rho\lambda\). Thus queue stability alone does not control unmanaged memory; SFMSN enforces~\eqref{eq:memory-budget}--\eqref{eq:memory-dynamics}.
\end{proof}

The local reason is that visible queues and nominal capacities may be identical while compatible memory differs.

\begin{proposition}[Minimax decision-state separation]
\label{prop:separation} 
Consider one unit of demand that must be assigned to exactly one of two branches with identical visible queues and nominal capacities, and let \(K>1\). In one instance branch 1 has service penalty \(1\) and branch 2 has penalty \(K\); in the paired instance the penalties are reversed because the resident-memory states are reversed. Any randomized rule whose split is invariant to these paired memory states has worst-case expected one-step penalty at least \((K+1)/2\), whereas a memory-aware rule pays \(1\). Equivalently, the worst-case additive gap is at least \((K-1)/2\). 
\end{proposition}
\begin{proof}[Proof sketch]
Let \(\alpha\) be the probability of assigning the unit to branch 1. The two expected penalties are
\(\alpha+(1-\alpha)K\) and
\(\alpha K+(1-\alpha)\). Their maximum is minimized at
\(\alpha=1/2\), where it equals \((K+1)/2\). A memory-aware rule observes which branch is compatible and assigns the unit to that branch, incurring penalty \(1\).
\end{proof}

For queue vector \(Q\), inherited memory \(y\), and one-slot control \(\gamma\), define
\begin{align}
\Phi(Q,y,\gamma)
={}&\sum_{(i,j)\in\E}\sum_{c\in\C}F_{ij}^c
   \bigl(Q_i^c-Q_j^c-\nu d_{ij}^c\bigr) \notag\\
&+\sum_{i\in\N}\sum_{a\in\A_i}S_i^a
   \bigl(Q_i^{\iota(a)}-\nu p_i(a,Y_i^\gamma)\bigr) \notag\\
&-\sum_{i\in\N}\sum_{a\in\A_i}S_i^a
  \sum_{c'\in\C}P_i(c'\mid a,Y_i^\gamma)Q_i^{c'} \\
&-\nu\bar h^{\rm mem}(y,\gamma)
\label{eq:mabp-weight}\\[-1mm]
={}&\sum_{(i,c)\in\mathcal I}Q_i^c g_i^c(y,\gamma)-\nu J(y,\gamma).
\label{eq:mabp-net-service-form}
\end{align}
The second form is used in the proofs: \(\Phi\) rewards expected net queue drainage and subtracts weighted routing, service, and memory penalties. A one-slot maximizer prices active memory but can undervalue retention whose benefit appears later.

Let \(H\) be a frame length, \(t_n=nH\), \(Q^n=Q(t_n)\), and \(Y_0=Y^-(t_n)\). For \(\tau=0,\ldots,H-1\), write \(Y_n(\tau)=Y^-(t_n+\tau)\) and \(\gamma_n(\tau)=\gamma(t_n+\tau)\). A causal frame policy uses only the realized queue--memory history inside the frame; the physical queues evolve normally, but the objective freezes \(Q^n\) as the linear weight. Define
\begin{equation}
\Phi_H(Q^n,Y_0,\pi)
=
\EE_\pi^{Q^n,Y_0}\!\left[
\sum_{\tau=0}^{H-1}
\Phi\bigl(Q^n,Y_n(\tau),\gamma_n(\tau)\bigr)
\right],
\label{eq:frame-mabp-objective}
\end{equation}
where the expectation covers arrivals, service outcomes, generated objects, and internal randomization.

\begin{definition}[Frame-\mabp]
\label{def:frame-mabp}
Exact frame-\mabp\ maximizes~\eqref{eq:frame-mabp-objective} over causal \(H\)-slot policies whose realized actions satisfy \(\gamma(t_n+\tau)\in\Gamma(Q(t_n+\tau),Y^-(t_n+\tau))\) on every sample path, executes the maximizer for the frame, and replans at \(t_{n+1}\). The case \(H=1\) is myopic; \(H>1\) values retained memory through later service operators.
\end{definition}

\subsection{Practical One-Slot Surrogate}

The exact frame oracle is analytic, so experiments use a one-slot surrogate: it preserves the instantaneous \(\Phi\) weights and approximates delayed retention by a frozen-queue local value. It is not automatically throughput-optimal; the guarantee requires the bounded-loss condition in Corollary~\ref{cor:additive-approx}.

Routing and service use
$
W_{ij}^c(t)=Q_i^c(t)-Q_j^c(t)-\nu d_{ij}^c,\qquad
W_i^a(q_i;Y_i)=q_i^{\iota(a)}
-\sum_{c'\in\C}P_i(c'\mid a,Y_i)q_i^{c'}
-\nu p_i(a,Y_i).
$
Let \(\zeta(t)=(F(t),u(t),R(t),Y^{\rm act}(t),S(t))\), and let \(\Gamma^{\rm pre}(Q(t),Y^-(t))\) collect the pre-service tuples satisfying migration-source, residency, memory-budget, link-capacity, compute-capacity, queue-feasibility, activation, and transfer-link constraints. The pre-service decision is
\begin{align}
\max_{\zeta(t)\in\Gamma^{\rm pre}(Q(t),Y^-(t))}
\quad
&\sum_{(i,j)\in\E}\sum_{c\in\C}
F_{ij}^c(t)W_{ij}^c(t) \notag\\
&+\sum_{i\in\N}\sum_{a\in\A_i}
S_i^a(t)W_i^a\bigl(Q_i(t);Y_i^{\rm act}(t)\bigr)\notag \\
&-\nu\sum_{(i,j)\in\E}\sum_{m\in\M}
\kappa_{ij}^m u_{ij}^m(t).
\label{eq:practical-pre-service}
\end{align}
Generated objects are unknown before service, so retention is handled by the post-service proxy below.

For resident set \(R_i\), let \(\mathcal Y_i^{\rm act}(R_i)\) be the feasible active subsets. Define the frozen-queue node value
\begin{align}
\Psi_i(R_i;q_i)
:=\max_{Y_i\in\mathcal Y_i^{\rm act}(R_i),\,s\in\ZZ_+^{|\A_i|}}
&\sum_{a\in\A_i}s_aW_i^a(q_i;Y_i) \notag
\label{eq:node-service-value}\\[-1mm]
\text{s.t.}\quad
&\sum_a s_aw_i(a,Y_i)\le\mu_i,\quad\\
&\sum_{a:\iota(a)=c}s_a\le q_i^c,\ (i,c)\in\mathcal I .
\nonumber
\end{align}
After service, the candidate set is \(\mathcal M_i^{\rm cand}(t)=R_i(t)\cup G_i(t)\). A precommitted retention rule is evaluated on the realized candidates using frozen \(Q_i(t)\):
\begin{equation}
\mathcal V_i(Z;t)
=
\Psi_i(Z;Q_i(t))
-\nu\theta_i\sum_{m\in Z}b_m
-\nu\sum_{m\in Z\cap G_i(t)}\omega_i^m,
\label{eq:retention-objective}
\end{equation}
over feasible \(Z\subseteq\mathcal M_i^{\rm cand}(t)\). Exact proxy retention maximizes~\eqref{eq:retention-objective}; a greedy implementation uses
\(\Delta_i^{\rm mem}(m\mid Z;t)=\mathcal V_i(Z\cup\{m\};t)-\mathcal V_i(Z;t)\).
This proxy estimates next-slot service value only, so its gap to the exact \(H\)-slot oracle must be uniformly bounded in queue magnitude to preserve the theorem-level guarantee.

\begin{algorithm}[t]
\caption{Exact Frame-Based MABP Oracle}
\label{alg:frame-mabp}
\begin{algorithmic}[1]
\Require \(H\), \(\nu\), arrival law, service-outcome kernels, and memory-transition kernels
\For{each frame boundary \(t_n=nH\)}
    \State Observe \(Q(t_n)\) and \(Y^-(t_n)\).
    \State Compute a causal feasible \(H\)-slot policy maximizing \(\Phi_H(Q(t_n),Y^-(t_n),\pi)\).
    \State Execute it for \(H\) slots, updating queues and memory by \eqref{eq:queue-dynamics} and \eqref{eq:memory-dynamics}.
\EndFor
\end{algorithmic}
\end{algorithm}

\begin{algorithm}[t]
\caption{Practical One-Slot MABP Surrogate}
\label{alg:practical-mabp}
\begin{algorithmic}[1]
\Require \(\nu\), primitive profiles or estimators, and queue/memory telemetry
\For{\(t=0,1,\ldots\)}
    \State Observe \(Q(t)\), \(Y^-(t)\), and feasible transfer candidates.
    \State Solve~\eqref{eq:practical-pre-service} for routing, transfer, activation, and service.
    \State Execute the pre-service decisions; observe successor jobs and generated objects \(G_i(t)\).
    \ForAll{nodes \(i\)}
        \State Set \(\mathcal M_i^{\rm cand}(t)=R_i(t)\cup G_i(t)\).
        \State Evaluate the precommitted rule~\eqref{eq:retention-objective}, or its greedy marginal version, to obtain feasible \(Z_i\).
        \State Set \(Y_i^-(t+1)=Z_i\).
    \EndFor
    \State Add exogenous arrivals and update \(Q(t+1)\) by~\eqref{eq:queue-dynamics}.
\EndFor
\end{algorithmic}
\end{algorithm}

\section{Stability and Cost Guarantees}
\label{sec:analysis}
This section converts the frozen queue-weight objective into a finite-frame Lyapunov bound, uses Assumption~\ref{ass:realization} as a slack comparator, and derives the penalty--backlog tradeoff.

\begin{assumption}[Bounded finite primitives]
\label{ass:bounded}
The commodity, action, memory-state, and one-slot feasible-control sets are finite. Arrivals have uniformly bounded conditional second moments, and routing amounts, service amounts, memory costs, and service penalties are uniformly bounded under every feasible control.
\end{assumption}
Throughout, let \(L(Q)=\frac12\sum_{(i,c)\in\mathcal I}(Q_i^c)^2\),
\(Q\cdot v=\sum_{(i,c)\in\mathcal I}Q_i^c v_i^c\), and
\(\EE_n[\cdot]=\EE[\cdot\mid Q(t_n),Y^-(t_n)]\). At \(t_n=nH\), define
\(\Delta_H(t_n)=\EE_n[L(Q(t_n+H))-L(Q(t_n))]\).

By Assumption~\ref{ass:bounded}, there is a finite \(B_H\), independent of \(Q(t_n)\) and \(Y^-(t_n)\), such that every causal \(H\)-slot policy satisfies
\begin{align}
\Delta_H(t_n)
\le{}&
B_H+H Q(t_n)\cdot\lambda \notag\\
&-\EE_n\!\left[
\sum_{\tau=0}^{H-1}\sum_{(i,c)\in\mathcal I}
Q_i^c(t_n)g_i^c(Y_n(\tau),\gamma_n(\tau))
\right].
\label{eq:frame-drift}
\end{align}
This is the only Lyapunov estimate used below; memory enters only through \(g_i^c\). Also,
\(\EE[h(t)\mid Y^-(t),\gamma(t)]=J(Y^-(t),\gamma(t))\).

\begin{theorem}[Throughput of exact frame-\mabp]
\label{thm:throughput}
Suppose Assumptions~\ref{ass:bounded} and~\ref{ass:realization} hold. Fix \(\nu\ge0\) and an arrival vector \(\lambda\) supported by a reachable occupation measure \(x\) with uniform slack \(\delta>0\). If \(H\ge H_x\) and \(H\delta>L_x\), then exact frame-\mabp\ strongly stabilizes all queues. Moreover, for some finite \(B_{H,x,\nu}\), independent of \(Q(t_n)\) and \(Y^-(t_n)\),
$
\EE_n\!\left[
L(Q(t_n+H))-L(Q(t_n))
+\nu\sum_{\tau=0}^{H-1}h(t_n+\tau)
\right]
\le
B_{H,x,\nu}
-\bigl(H\delta-L_x\bigr)\|Q(t_n)\|_1 .
$
\end{theorem}

\begin{proof}[Proof sketch]
Adding the conditional expected penalty to~\eqref{eq:frame-drift} and using
\(\Phi(Q,y,\gamma)=Q\cdot g(y,\gamma)-\nu J(y,\gamma)\) gives
$
\Delta_H(t_n)
+\nu\EE_n\!\left[\sum_{\tau=0}^{H-1}h(t_n+\tau)\right]
\le
B_H+H Q(t_n)\cdot\lambda
-\Phi_H(Q(t_n),Y^-(t_n),\pi_{\mabp}).
$
Let \(\pi_x\) be the queue-compatible realization of \(x\). Exact optimality and~\eqref{eq:uniform-realization}--\eqref{eq:penalty-realization} give
$
\Phi_H(Q(t_n),Y^-(t_n),\pi_{\mabp})
\ge
H Q(t_n)\cdot\bar g(x)
-L_x\|Q(t_n)\|_1
-C_{H,x}
-\nu(H\bar J(x)+D_{H,x}).
$
Since \(\bar g_i^c(x)\ge\lambda_i^c+\delta\), substitution yields the stated drift bound after absorbing constants into \(B_{H,x,\nu}\). Because \(H\delta>L_x\) and \(h(t)\) is bounded below, the frame drift is negative outside a bounded set. The standard frame Lyapunov criterion gives strong stability at frame boundaries, and bounded intra-frame increments extend it to all slots.
\end{proof}

To quantify the penalty--backlog tradeoff, compare exact frame-\mabp\ with the best realizable slack-\(\delta\) occupation measure. For any policy \(\pi\), define
$
\bar J^\pi=\limsup_{T\to\infty}\frac1T\sum_{t<T}\EE[h^\pi(t)],
\quad
\bar Q^\pi=\limsup_{T\to\infty}\frac1T\sum_{t<T}\EE[\|Q^\pi(t)\|_1].
$
For fixed \(\delta>0\), let \(J_\delta^\star\) be the infimum of \(\bar J(x)\) over reachable occupation measures that support \(\lambda\) with slack at least \(\delta\) and satisfy Assumption~\ref{ass:realization}.

\begin{theorem}[Frame drift-plus-penalty]
\label{thm:dpp}
Suppose \(\nu>0\), and assume that the infimum defining \(J_\delta^\star\) is attained by a reachable occupation measure \(x_\delta^\star\) satisfying Assumption~\ref{ass:realization}. Let \(H_\delta\), \(L_\delta\), \(C_{H,\delta}\), and \(D_{H,\delta}\) be the realization constants associated with \(x_\delta^\star\). For every fixed \(H\ge H_\delta\) satisfying \(H\delta>L_\delta\), exact frame-\mabp\ satisfies
\begin{align}
\bar J^{\mabp}
&\le
J_\delta^\star
+\frac{D_{H,\delta}}{H}
+\frac{B_H+C_{H,\delta}}{\nu H},
\notag\\
\bar Q^{\mabp}
&=O(\nu)+O(1),
\end{align}
where \(B_H\) is the constant in~\eqref{eq:frame-drift}. The hidden constants may depend on \(H\), \(\delta\), \(L_\delta\), \(J_{\min}\), and the primitive bounds, but not on \(\nu\).
\end{theorem}

\begin{proof}[Proof sketch]
Repeating the comparison in Theorem~\ref{thm:throughput} with \(x_\delta^\star\) gives
$
\Delta_H(t_n)
+\nu\EE_n\!\left[\sum_{\tau=0}^{H-1}h(t_n+\tau)\right]
\le
B_H+C_{H,\delta}
+\nu(HJ_\delta^\star+D_{H,\delta})
-\bigl(H\delta-L_\delta\bigr)\|Q(t_n)\|_1 .
$
Summing over frames, telescoping \(L(Q)\), dropping the nonpositive backlog term, dividing by \(\nu H\), and taking \(T\to\infty\) gives the penalty bound. For backlog, retain the negative backlog term and use \(h(t)\ge J_{\min}\), which yields
$
\limsup_{N\to\infty}
\frac1N\sum_{n<N}\EE[\|Q(t_n)\|_1]
\le
\frac{
B_H+C_{H,\delta}
+\nu\bigl(H(J_\delta^\star-J_{\min})+D_{H,\delta}\bigr)
}{
H\delta-L_\delta
}.
$
Thus the frame-boundary backlog is \(O(\nu)+O(1)\), and bounded intra-frame increments extend the same scaling to \(\bar Q^{\mabp}\).
\end{proof}

\begin{corollary}[Additive frame approximation] 
\label{cor:additive-approx} 
Fix \(\nu\). Suppose that, for every reachable \(Q,Y\), an approximate \(H\)-slot solver satisfies
$
\Phi_H(Q,Y,\widetilde\pi)
\ge
\sup_\pi\Phi_H(Q,Y,\pi)-C_H^{\rm app},
$
where the supremum is over causal, physically feasible frame policies and \(C_H^{\rm app}\) is finite and uniform in \(Q,Y\). Then the drift inequalities in Theorems~\ref{thm:throughput} and~\ref{thm:dpp} only gain the queue-independent term \(C_H^{\rm app}\). Hence throughput is preserved whenever \(H\delta>L_x\), and
$
\bar J^{\widetilde\pi}
\le
J_\delta^\star
+\frac{D_{H,\delta}}{H}
+\frac{B_H+C_{H,\delta}+C_H^{\rm app}}{\nu H}.
$
\end{corollary}
\begin{proof}[Proof sketch]
Replace the exact optimality step
\(\Phi_H(\pi_{\mabp})\ge \Phi_H(\pi_x)\)
by the additive inequality above. The loss \(C_H^{\rm app}\) does not scale with \(\|Q\|_1\), so it only changes the constant term and leaves the negative backlog coefficient unchanged.
\end{proof}

\begin{remark}[Scope of the practical rule]
Algorithm~\ref{alg:practical-mabp} is not automatically throughput-optimal. It inherits the frame theorem only when its accumulated \(\Phi_H\)-loss is uniformly bounded in queue magnitude; generic greedy retention or errors in \(P_i\) need not satisfy this condition.
\end{remark}

\section{Evaluation}
\label{sec:evaluation}

We evaluate Algorithm~\ref{alg:practical-mabp}, not the exact frame oracle. Unless stated otherwise, \mabp\ uses greedy feasible max-weight decisions in~\eqref{eq:practical-pre-service} and marginal-value retention in~\eqref{eq:retention-objective}. Oracle-Sur applies near-exact local search with known primitives, measuring the one-slot optimization gap rather than a capacity upper bound.

\subsection{Methodology}

\paragraph{Workloads and parameters.}
The simulator instantiates finite RAG, MoE, and agentic SFMSNs whose commodities and objects follow Section~\ref{sec:model}; it is not a trace replay of \texttt{Qwen/Qwen2.5-7B-Instruct}. We use a two-server separation case, a synthetic occupation-boundary network, and a mixed RAG--MoE--agentic graph. Normalized load \(\rho\) scales Poisson arrivals, and \(B_i(\beta)=\beta B_i^{\max}\), where \(B_i^{\max}\) is the eligible-object footprint. Defaults are horizon \(600\), \(\rho_0=0.85\), \(\beta=0.25\), reuse skew \(1.0\), transition noise \(0.10\), migration cost \(2.0\), noncached factor \(K=4\), and \(\nu=5\). Sweeps use \(\rho\in\{0.1,\ldots,1.2\}\), \(\beta\in\{0.05,0.10,0.20,0.30,0.40,0.50\}\), \(K\in\{1,2,4,8,16\}\), \(\nu\in\{0,1,2,5,10,20\}\), and Zipf skew in \(\{0.5,1.0,1.5,2.0\}\).

\paragraph{Policies and baselines.}
All policies share the same feasibility constraints and realized primitives; only ranking scores differ. SP uses shortest-path cost, JSQ the smallest eligible queue, BP memory-oblivious differentials, Cost-BP an added reference cost, and GreedyCache-BP popularity-based retention. GNN uses the same telemetry as \mabp{} before the common feasibility projection. Memory-oblivious scores ignore \(Y_i\) in ranking but obey state-dependent feasibility and may obtain passive hits; their successor law is averaged over profiled memory states. Learn-\mabp\ uses frozen estimates of \(w_i\), \(p_i\), and \(P_i\), with unseen pairs using no-reuse primitives.

\paragraph{Metrics.}
We discard the first \(20\%\) of slots and report mean \(\pm\) standard deviation. Stability requires post-warm-up backlog slope below \(0.035\) and mean backlog below \(120\); stable load is the largest tested load passing both. Fixed-load metrics at \(\rho_0=0.85\) may place weaker baselines outside their stable region. Penalty is per served request, P95 delay uses completed requests, hit ratio counts compatible memory-assisted actions, churn counts insertions/evictions, and completion ratio is served work divided by arrivals.

\subsection{Simulation Results}

\begin{table}[t]
\centering
\caption{Mixed-workflow results. Stable load is the largest stable tested load; other metrics are measured at \(\rho_0=0.85\). Entries report mean \(\pm\) standard deviation over repeated runs. Higher is better for stable load and hit ratio; lower is better otherwise.}
\label{tab:main-results}
\scriptsize
\setlength{\tabcolsep}{3pt}
\resizebox{\linewidth}{!}{%
\begin{tabular}{lccccc}
\toprule
Policy & Stable load & Penalty & P95 delay & Hit ratio & Churn \\
\midrule
SP              & $0.5$ & $3.86{\pm}0.00$ & $2249.03{\pm}187.23$ & $0.05{\pm}0.00$ & $0.06{\pm}0.00$ \\
JSQ             & $0.5$ & $3.85{\pm}0.00$ & $1688.81{\pm}195.79$ & $0.06{\pm}0.00$ & $0.05{\pm}0.00$ \\
BP              & $0.6$ & $3.86{\pm}0.00$ & $1295.90{\pm}98.58$  & $0.06{\pm}0.00$ & $0.06{\pm}0.00$ \\
Cost-BP         & $0.6$ & $3.69{\pm}0.00$ & $1219.89{\pm}98.19$  & $0.10{\pm}0.00$ & $0.06{\pm}0.00$ \\
GreedyCache-BP  & $0.7$ & $3.52{\pm}0.00$ & $932.02{\pm}56.95$   & $0.32{\pm}0.00$ & $0.07{\pm}0.00$ \\
GNN scheduler   & $0.8$ & $3.41{\pm}0.00$ & $562.21{\pm}20.22$   & $0.33{\pm}0.00$ & $0.06{\pm}0.00$ \\
\mabp           & $0.9$ & $3.12{\pm}0.00$ & $129.13{\pm}16.91$   & $0.43{\pm}0.00$ & $0.04{\pm}0.00$ \\
Oracle-Sur      & $1.0$ & $3.05{\pm}0.00$ & $88.76{\pm}32.27$    & $0.46{\pm}0.00$ & $0.04{\pm}0.00$ \\
\bottomrule
\end{tabular}}
\vspace{-2mm}
\end{table}

\begin{table*}[t]
\centering
\caption{Real multi-GPU LLM-system results. Entries report mean $\pm$ the two-sided 95\% Student-\(t\) confidence-interval half-width across independent runs; bold marks the best point estimate.}
\label{tab:real-system}
\scriptsize
\begin{tabular}{lccccc}
\toprule
Metric & JSQ & BP & Aff. & GNN-Sched. & MABP-Route \\
\midrule
Throughput (req/s) & 6.05$\pm$0.10 & 6.04$\pm$0.07 & 6.14$\pm$0.06 & 6.10$\pm$0.10 & \textbf{6.21$\pm$0.08} \\
SLO goodput (req/s) & 2.25$\pm$0.15 & 2.41$\pm$0.17 & 2.61$\pm$0.19 & 2.64$\pm$0.22 & \textbf{2.82$\pm$0.33} \\
P95 TTFT (ms) & 71.82$\pm$2.40 & 71.62$\pm$1.49 & 73.37$\pm$2.02 & 71.77$\pm$2.08 & \textbf{70.53$\pm$0.86} \\
P95 E2E latency (s) & 2.05$\pm$0.18 & 2.18$\pm$0.03 & 1.94$\pm$0.06 & 2.08$\pm$0.14 & \textbf{1.93$\pm$0.14} \\
GPU seconds/request & 0.994$\pm$0.006 & 0.993$\pm$0.003 & 0.985$\pm$0.007 & 0.986$\pm$0.003 & \textbf{0.965$\pm$0.003} \\
KV hit ratio & 0.884$\pm$0.023 & 0.884$\pm$0.007 & 0.972$\pm$0.006 & 0.908$\pm$0.010 & \textbf{0.998$\pm$0.014} \\
Downstream jobs/request & 1.278$\pm$0.066 & 1.268$\pm$0.025 & 1.179$\pm$0.015 & 1.196$\pm$0.032 & \textbf{1.056$\pm$0.051} \\
Drop rate (\%) & $0.00{\pm}0.00$ & $0.00{\pm}0.00$ & $0.00{\pm}0.00$ & $0.00{\pm}0.00$ & $0.00{\pm}0.00$ \\
P95 scheduler time ($\mu$s) & 38.64$\pm$4.31 & \textbf{31.68$\pm$6.18} & 42.93$\pm$9.03 & 48.25$\pm$12.61 & 33.74$\pm$12.06 \\
Mean attempts & $1.00{\pm}0.00$ & $1.00{\pm}0.00$ & $1.00{\pm}0.00$ & $1.00{\pm}0.00$ & $1.00{\pm}0.00$ \\
\bottomrule
\end{tabular}
\end{table*}

Table~\ref{tab:main-results} separates stability from fixed-load stress performance. \mabp\ reaches stable load \(0.9\), exceeding GNN scheduler, GreedyCache-BP, and BP by \(0.1\), \(0.2\), and \(0.3\), respectively. At \(\rho_0=0.85\), it reduces P95 delay by \(90.0\%\) over BP and \(77.0\%\) over GNN scheduler, reduces penalty by \(19.2\%\) and \(8.5\%\), respectively, and stays within \(2.3\%\) of Oracle-Sur. The higher hit ratio with lower churn supports selective state reuse rather than aggressive replacement.

\begin{figure}[t]
\centering
\begin{minipage}[t]{0.49\linewidth}
\centering
\includegraphics[width=\linewidth]{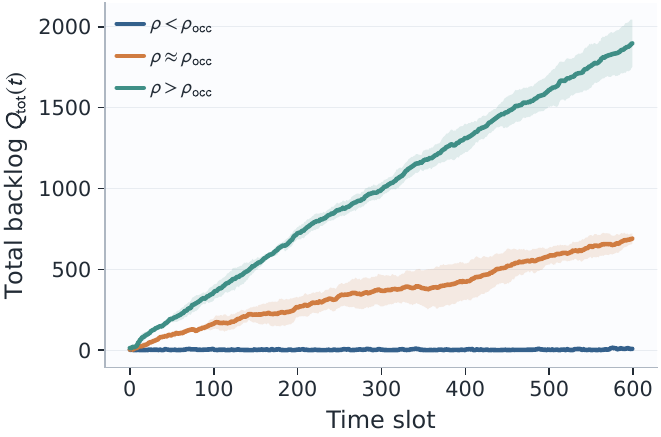}\\[-1mm]
{\scriptsize (a) Backlog trace below/near/above boundary}
\end{minipage}\hfill
\begin{minipage}[t]{0.49\linewidth}
\centering
\includegraphics[width=\linewidth]{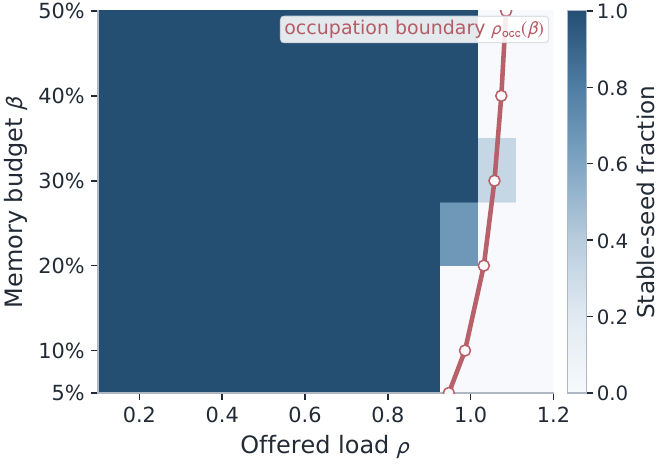}\\[-1mm]
{\scriptsize (b) Occupation boundary versus memory budget}
\end{minipage}
\caption{Stability and capacity-region diagnostics. (a) The below-boundary load keeps the backlog controlled, while near-boundary and above-boundary loads produce visibly larger backlog growth. (b) The occupation-measure outer boundary expands as the memory budget increases.}
\label{fig:stability-capacity}
\vspace{-2mm}
\end{figure}

Fig.~\ref{fig:stability-capacity} validates the stability mechanism. The controlled trace separates below-, near-, and above-boundary regimes at loads \(0.66\), \(0.96\), and \(1.14\), with only the below-boundary case keeping backlog controlled. The occupation boundary rises from \(\rho_{\rm occ}=0.947\) at \(\beta=0.05\) to \(1.086\) at \(\beta=0.50\), and the empirical stable grid load reaches \(1.0\) once \(\beta\ge0.20\).

\begin{figure}[t]
\centering
\begin{minipage}[t]{0.49\linewidth}
\centering
\includegraphics[width=\linewidth]{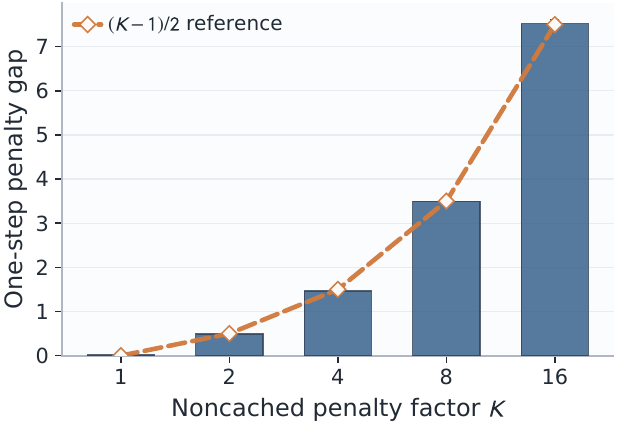}\\[-1mm]
{\scriptsize (a) Memory-oblivious separation versus \(K\)}
\end{minipage}\hfill
\begin{minipage}[t]{0.49\linewidth}
\centering
\includegraphics[width=\linewidth]{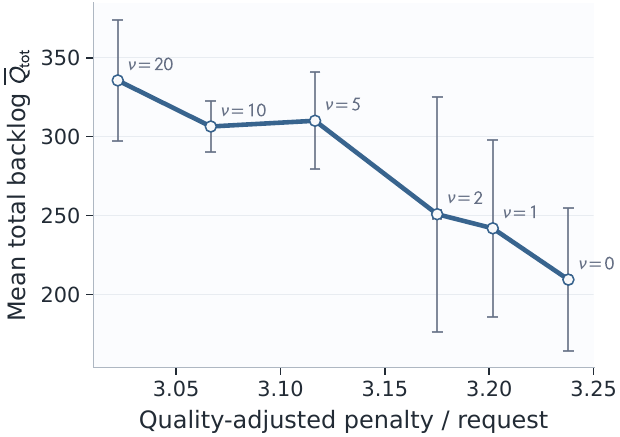}\\[-1mm]
{\scriptsize (b) Penalty--backlog tradeoff versus \(\nu\)}
\end{minipage}
\caption{Mechanism-level tradeoffs. (a) The one-step BP--\mabp\ penalty gap grows with the noncached penalty factor \(K\). (b) Increasing \(\nu\) moves the operating point toward lower penalty and larger backlog.}
\label{fig:separation-tradeoff}
\vspace{-2mm}
\end{figure}

Fig.~\ref{fig:separation-tradeoff} tests the mechanism-level predictions. In the paired separation instance of Proposition~\ref{prop:separation}, the BP-minus-\mabp\ one-step penalty gap is near zero at \(K=1\) and grows to \(1.46\), \(3.50\), and \(7.52\) at \(K=4,8,16\), matching the scaling toward \((K-1)/2=7.5\). Increasing \(\nu\) from \(0\) to \(20\) lowers penalty from \(3.24\) to \(3.02\), while mean backlog increases from \(209.4\) to \(335.6\), as expected for the DPP tradeoff.

\begin{figure}[t]
\centering
\begin{minipage}[t]{0.49\linewidth}
\centering
\includegraphics[width=\linewidth]{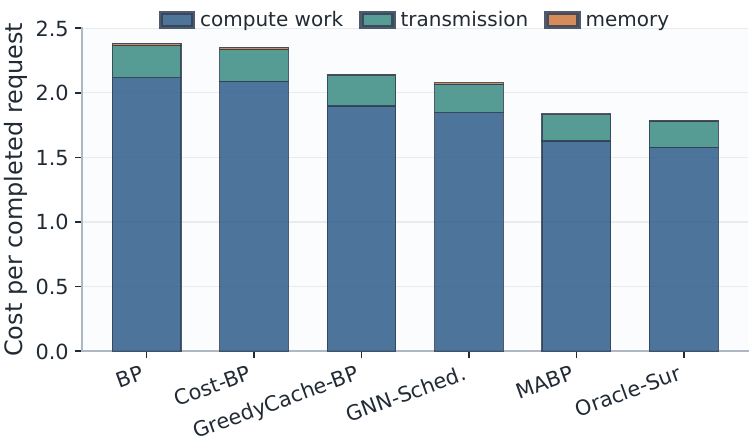}\\[-1mm]
{\scriptsize (a) Penalty-component breakdown}
\end{minipage}\hfill
\begin{minipage}[t]{0.49\linewidth}
\centering
\includegraphics[width=\linewidth]{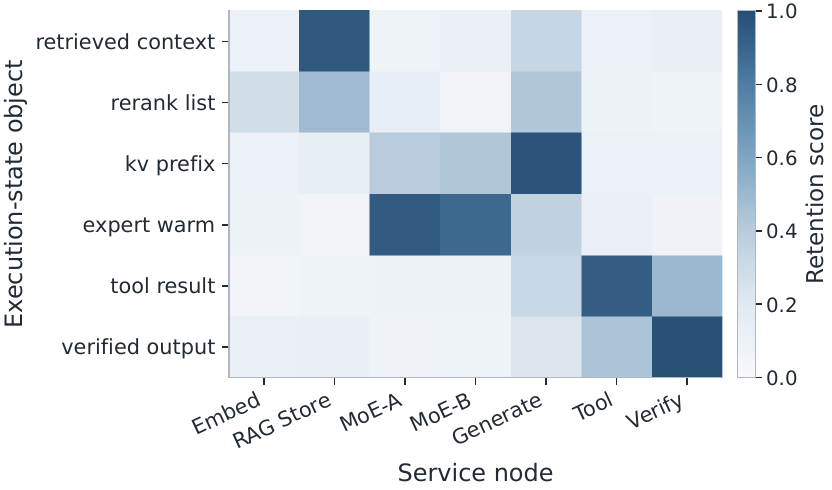}\\[-1mm]
{\scriptsize (b) Learned memory-placement heatmap}
\end{minipage}
\caption{Cost composition and memory placement. (a) \mabp\ reduces compute work, transmission cost, and memory cost on the mixed workload. (b) Retention scores concentrate on semantically compatible node--object pairs rather than uniformly caching all objects.}
\label{fig:cost-cache}
\vspace{-2mm}
\end{figure}

Fig.~\ref{fig:cost-cache} shows where the penalty gain comes from: \mabp\ reduces compute work from \(2.116\) under BP and \(1.846\) under GNN scheduler to \(1.628\), lowers transmission cost from \(0.251\) to \(0.204\), and concentrates retention on semantically compatible node--object pairs.

\begin{figure}[t]
\centering
\begin{minipage}[t]{0.49\linewidth}
\centering
\includegraphics[width=\linewidth]{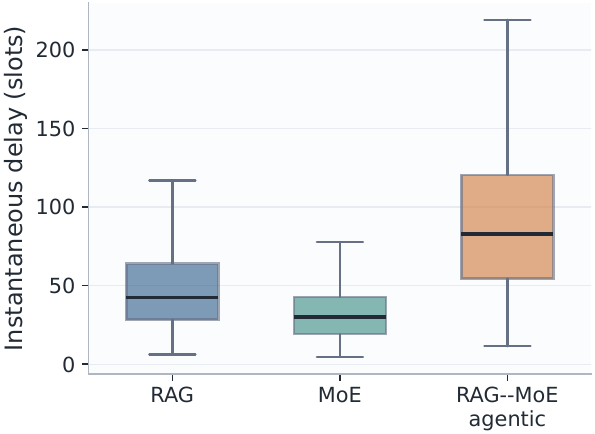}\\[-1mm]
{\scriptsize (a) Tail-delay distribution by workload}
\end{minipage}\hfill
\begin{minipage}[t]{0.49\linewidth}
\centering
\includegraphics[width=\linewidth]{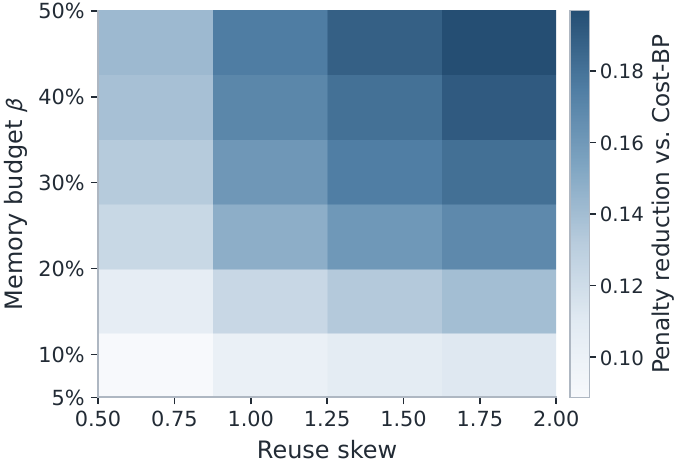}\\[-1mm]
{\scriptsize (b) Memory-budget/reuse-skew sensitivity}
\end{minipage}
\caption{Tail behavior and sensitivity. (a) The mixed workflow has the heaviest tail because it combines RAG, MoE, and agentic transition stages. (b) The \mabp{} gain grows as either memory budget or reuse skew increases.}
\label{fig:tail-sensitivity}
\vspace{-2mm}
\end{figure}

Fig.~\ref{fig:tail-sensitivity} confirms that the mixed workflow is the hardest tail setting, with P95 delay \(205.18\) versus \(113.04\) for RAG and \(69.96\) for MoE. The penalty gain over Cost-BP increases with both memory budget and reuse skew, from \(10.2\%\) to \(17.6\%\) across \(\beta\) and from \(12.2\%\) to \(16.5\%\) across skew.

\begin{figure}[t]
\centering
\includegraphics[width=0.68\linewidth]{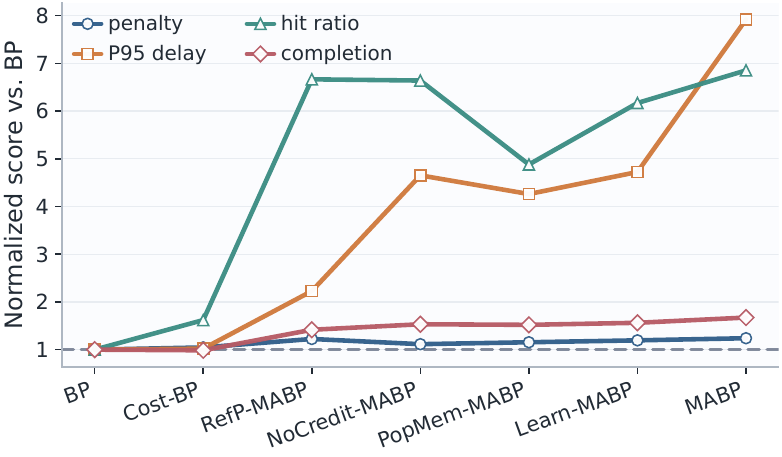}
\caption{Component ablations normalized to BP. For lower-is-better metrics the score is \(x_{\rm BP}/x\), and for higher-is-better metrics the score is \(x/x_{\rm BP}\).}
\label{fig:ablation-profile}
\vspace{-2mm}
\end{figure}

Fig.~\ref{fig:ablation-profile} isolates the practical policy terms at \(\rho=0.86\), \(\beta=0.25\), \(K=4\), and \(\nu=5\). Removing downstream transition pressure, reuse credit, or marginal-value retention raises P95 delay to \(587.45\), \(281.25\), and \(307.10\), respectively; Learn-\mabp\ gives \(277.04\), while full \mabp\ gives \(165.26\) and the lowest penalty \(3.12\). These terms are therefore complementary.

\subsection{Real Multi-GPU LLM-System Evaluation}
\label{sec:real-system}

We also evaluate a routing-only deployment, \texttt{MABP-Route}, in a real four-GPU vLLM system serving \texttt{Qwen/Qwen2.5-7B-Instruct} through an OpenAI-compatible chat interface. An external asynchronous dispatcher routes a tool-heavy RAG, MoE-style, and agentic tool/verification workload with reusable prefix/KV and downstream tool-result keys. Because vLLM exposes prefix/KV reuse but not full object-level eviction, \texttt{MABP-Route} controls routing while leaving low-level KV eviction to the engine.

All policies share the same request stream, model, worker pool, concurrency limit, and timeout configuration: inter-arrival time \(0.13\) s, at most \(12\) in-flight requests, timeout \(180\) s, one retry, temperature \(0\), maximum output length \(72\), and SLO \(1.7\) s. JSQ uses in-flight queue length; BP uses queue pressure with a completion tie-breaker; Aff. favors prefix locality; GNN-Sched. uses the same telemetry as \mabp{} with a learned graph-score rule; \texttt{MABP-Route} combines queue pressure, prefix/KV reuse, downstream tool-state reuse, and overload damping.

Table~\ref{tab:real-system} shows that \texttt{MABP-Route} attains the best point estimates for SLO goodput, E2E latency, GPU work, KV hit ratio, and downstream jobs, with zero drops and one attempt per request for all policies. Relative to GNN-Sched., it improves SLO goodput by \(6.8\%\), reduces P95 E2E latency by \(7.2\%\), reduces GPU seconds/request by \(2.1\%\), raises KV hit ratio from \(0.908\) to \(0.998\), and reduces downstream jobs/request by \(11.7\%\). The real-system margins are smaller than in simulation because all policies share the same vLLM cache manager and safe concurrency limit, and some confidence intervals overlap; we therefore interpret them as mechanism-consistent gains rather than broad statistical dominance.

\section{Conclusion}

This paper develops \mabp{} for stateful foundation-model service networks, where reusable execution states affect not only local work and penalty but also downstream successor generation. By modeling memory as a controlled network state, we derive an occupation-measure capacity outer bound, identify when memory enlarges the stabilizable region, prove memory-aware/memory-oblivious separation, and give throughput and drift-plus-penalty guarantees for exact frame-\mabp{} with bounded-loss approximate extensions. Experiments on mixed RAG--MoE--agentic workloads and a real multi-GPU LLM serving system show that the practical one-slot surrogate improves stability, tail delay, state reuse, and SLO goodput.

\nocite{*}
\bibliographystyle{IEEEtran}
\bibliography{refs}

@inproceedings{tassiulas1990stability,
  title={Stability properties of constrained queueing systems and scheduling policies for maximum throughput in multihop radio networks},
  author={Tassiulas, Leandros and Ephremides, Anthony},
  booktitle={29th IEEE Conference on Decision and Control},
  pages={2130--2132},
  year={1990},
  organization={IEEE}
}

@book{neely2010stochastic,
  title={Stochastic network optimization with application to communication and queueing systems},
  author={Neely, Michael},
  year={2010},
  publisher={Morgan \& Claypool Publishers}
}

@book{georgiadis2006resource,
  title={Resource allocation and cross-layer control in wireless networks},
  author={Georgiadis, Leonidas and Neely, Michael J and Tassiulas, Leandros},
  year={2006},
  publisher={Now Publishers Inc}
}

@article{stolyar2004maxweight,
  title={Maxweight scheduling in a generalized switch: State space collapse and workload minimization in heavy traffic},
  author={Stolyar, Alexander L},
  journal={The Annals of Applied Probability},
  volume={14},
  number={1},
  pages={1--53},
  year={2004},
  publisher={Institute of Mathematical Statistics}
}

@article{dai2005maximum,
  title={Maximum pressure policies in stochastic processing networks},
  author={Dai, Jim G and Lin, Wuqin},
  journal={Operations Research},
  volume={53},
  number={2},
  pages={197--218},
  year={2005},
  publisher={INFORMS}
}

@inproceedings{neely2010universal,
  title={Universal scheduling for networks with arbitrary traffic, channels, and mobility},
  author={Neely, Michael J},
  booktitle={49th IEEE Conference on Decision and Control (CDC)},
  pages={1822--1829},
  year={2010},
  organization={IEEE}
}

@book{meyn2012markov,
  title={Markov chains and stochastic stability},
  author={Meyn, Sean P and Tweedie, Richard L},
  year={2012},
  publisher={Springer Science \& Business Media}
}

@techreport{quinn2015problem,
  title={Problem statement for service function chaining},
  author={Quinn, Paul and Nadeau, Tom},
  year={2015}
}

@article{zhang2024distributed,
  title={Distributed age-of-information scheduling with noma via deep reinforcement learning},
  author={Zhang, Congwei and Zou, Yifei and Zhang, Zuyuan and Yu, Dongxiao and G{\'o}mez, Jorge Torres and Lan, Tian and Dressler, Falko and Cheng, Xiuzhen},
  journal={IEEE Transactions on Mobile Computing},
  volume={24},
  number={1},
  pages={30--44},
  year={2024},
  publisher={IEEE}
}

@article{zou2024distributed,
  title={A distributed abstract mac layer for cooperative learning on internet of vehicles},
  author={Zou, Yifei and Zhang, Zuyuan and Zhang, Congwei and Zheng, Yanwei and Yu, Dongxiao and Yu, Jiguo},
  journal={IEEE Transactions on Intelligent Transportation Systems},
  volume={25},
  number={8},
  pages={8972--8983},
  year={2024},
  publisher={IEEE}
}

@inproceedings{zhang2025network,
  title={Network diffuser for placing-scheduling service function chains with inverse demonstration},
  author={Zhang, Zuyuan and Aggarwal, Vaneet and Lan, Tian},
  booktitle={IEEE INFOCOM 2025-IEEE Conference on Computer Communications},
  pages={1--10},
  year={2025},
  organization={IEEE}
}

@inproceedings{qiao2024br,
  title={Br-defedrl: Byzantine-robust decentralized federated reinforcement learning with fast convergence and communication efficiency},
  author={Qiao, Jing and Zhang, Zuyuan and Yue, Sheng and Yuan, Yuan and Cai, Zhipeng and Zhang, Xiao and Ren, Ju and Yu, Dongxiao},
  booktitle={Ieee infocom 2024-ieee conference on computer communications},
  pages={141--150},
  year={2024},
  organization={IEEE}
}

@inproceedings{yu2024look,
  title={Look-ahead robust network optimization with generative state predictions},
  author={Yu, Fei Xu and Zhang, Zuyuan and Grob, Emily and Adam, Gina and Coffey, Sean and Bastian, Nathaniel D and Lan, Tian},
  booktitle={AAAI 2025 Workshop on Artificial Intelligence for Wireless Communications and Networking (AI4WCN)},
  year={2024}
}

@inproceedings{zhang2026lisfc,
  title={Lisfc-search: Lifelong search for network sfc optimization under non-stationary drifts},
  author={Zhang, Zuyuan and Aggarwal, Vaneet and Lan, Tian},
  booktitle={IEEE INFOCOM 2026-IEEE Conference on Computer Communications},
  pages={1--6},
  year={2026},
  organization={IEEE}
}

@article{zhang2026counterfactual,
  title={Counterfactual Regret Minimization-Mixing for Noncooperative Stochastic Spectrum Games with Imperfect Information},
  author={Zhang, Zuyuan and Liu, Lingjia and Bastian, Nathaniel D and Lan, Tian},
  journal={IEEE Transactions on Networking},
  year={2026},
  publisher={IEEE}
}

@article{yu2026interactive,
  title={Interactive Critique-Revision Training for Reliable Structured LLM Generation},
  author={Yu, Fei Xu and Zhang, Zuyuan and Imani, Mahdi and Bastian, Nathaniel D and Lan, Tian},
  journal={arXiv preprint arXiv:2605.08327},
  year={2026}
}

@article{jing2024byzantine,
  title={Byzantine fault tolerant consensus in open wireless networks via an abstract mac layer},
  author={Jing, Guanlin and Zou, Yifei and Zhang, Zuyuan and Yu, Dongxiao and Dressler, Falko and Cheng, Xiuzhen},
  journal={IEEE Transactions on Communications},
  volume={73},
  number={3},
  pages={1909--1924},
  year={2024},
  publisher={IEEE}
}

@article{yang2025jamming,
  title={Jamming-resilient physical-to-virtual communications in digital twin edge networks},
  author={Yang, Li and Zou, Yifei and Zhang, Zuyuan and Wang, Peng and Yu, Dongxiao and Zubow, Anatolij and Dressler, Falko and Cheng, Xiuzhen},
  journal={IEEE Transactions on Networking},
  volume={33},
  number={4},
  pages={1731--1745},
  year={2025},
  publisher={IEEE}
}

@article{yang2026dig,
  title={DIG to Heal: Scaling General-purpose Agent Collaboration via Explainable Dynamic Decision Paths},
  author={Yang, Hanqing and Lee, Hyungwoo and Yao, Yuhang and Liu, Zhiwei and Liu, Kay and Chen, Jingdi and Joe-Wong, Carlee},
  journal={arXiv preprint arXiv:2603.00309},
  year={2026}
}

@article{bhamare2016survey,
  title={A survey on service function chaining},
  author={Bhamare, Deval and Jain, Raj and Samaka, Mohammed and Erbad, Aiman},
  journal={Journal of Network and Computer Applications},
  volume={75},
  pages={138--155},
  year={2016},
  publisher={Elsevier}
}

@article{kamran2021deco,
  title={DECO: Joint computation scheduling, caching, and communication in data-intensive computing networks},
  author={Kamran, Khashayar and Yeh, Edmund and Ma, Qian},
  journal={IEEE/ACM Transactions on Networking},
  volume={30},
  number={3},
  pages={1058--1072},
  year={2021},
  publisher={IEEE}
}

@article{mao2017survey,
  title={A survey on mobile edge computing: The communication perspective},
  author={Mao, Yuyi and You, Changsheng and Zhang, Jun and Huang, Kaibin and Letaief, Khaled B},
  journal={IEEE communications surveys \& tutorials},
  volume={19},
  number={4},
  pages={2322--2358},
  year={2017},
  publisher={IEEE}
}

@article{ndikumana2019joint,
  title={Joint communication, computation, caching, and control in big data multi-access edge computing},
  author={Ndikumana, Anselme and Tran, Nguyen H and Ho, Tai Manh and Han, Zhu and Saad, Walid and Niyato, Dusit and Hong, Choong Seon},
  journal={IEEE Transactions on mobile Computing},
  volume={19},
  number={6},
  pages={1359--1374},
  year={2019},
  publisher={IEEE}
}

@article{maia2024survey,
  title={A survey on integrated computing, caching, and communication in the cloud-to-edge continuum},
  author={Maia, Adyson and Boutouchent, Akram and Kardjadja, Youcef and Gherari, Manel and Soyak, Ece Gelal and Saqib, Muhammad and Boussekar, Kacem and Cilbir, Idil and Habibi, Sama and Ali, Soukaina Ouledsidi and others},
  journal={Computer Communications},
  volume={219},
  pages={128--152},
  year={2024},
  publisher={Elsevier}
}

@article{bommasani2021opportunities,
  title={On the opportunities and risks of foundation models},
  author={Bommasani, Rishi and Hudson, Drew A and Adeli, Ehsan and Altman, Russ and Arora, Simran and von Arx, Sydney and Bernstein, Michael S and Bohg, Jeannette and Bosselut, Antoine and Brunskill, Emma and others},
  journal={arXiv preprint arXiv:2108.07258},
  year={2021}
}

@article{vaswani2017attention,
  title={Attention is all you need},
  author={Vaswani, Ashish and Shazeer, Noam and Parmar, Niki and Uszkoreit, Jakob and Jones, Llion and Gomez, Aidan N and Kaiser, {\L}ukasz and Polosukhin, Illia},
  journal={Advances in neural information processing systems},
  volume={30},
  year={2017}
}

@article{lewis2020retrieval,
  title={Retrieval-augmented generation for knowledge-intensive nlp tasks},
  author={Lewis, Patrick and Perez, Ethan and Piktus, Aleksandra and Petroni, Fabio and Karpukhin, Vladimir and Goyal, Naman and K{\"u}ttler, Heinrich and Lewis, Mike and Yih, Wen-tau and Rockt{\"a}schel, Tim and others},
  journal={Advances in neural information processing systems},
  volume={33},
  pages={9459--9474},
  year={2020}
}

@article{shazeer2017outrageously,
  title={Outrageously large neural networks: The sparsely-gated mixture-of-experts layer},
  author={Shazeer, Noam and Mirhoseini, Azalia and Maziarz, Krzysztof and Davis, Andy and Le, Quoc and Hinton, Geoffrey and Dean, Jeff},
  journal={arXiv preprint arXiv:1701.06538},
  year={2017}
}

@article{fedus2022switch,
  title={Switch transformers: Scaling to trillion parameter models with simple and efficient sparsity},
  author={Fedus, William and Zoph, Barret and Shazeer, Noam},
  journal={Journal of Machine Learning Research},
  volume={23},
  number={120},
  pages={1--39},
  year={2022}
}

@inproceedings{yu2022orca,
  title={Orca: A distributed serving system for $\{$Transformer-Based$\}$ generative models},
  author={Yu, Gyeong-In and Jeong, Joo Seong and Kim, Geon-Woo and Kim, Soojeong and Chun, Byung-Gon},
  booktitle={16th USENIX symposium on operating systems design and implementation (OSDI 22)},
  pages={521--538},
  year={2022}
}

@inproceedings{li2023alpaserve,
  title={$\{$AlpaServe$\}$: Statistical multiplexing with model parallelism for deep learning serving},
  author={Li, Zhuohan and Zheng, Lianmin and Zhong, Yinmin and Liu, Vincent and Sheng, Ying and Jin, Xin and Huang, Yanping and Chen, Zhifeng and Zhang, Hao and Gonzalez, Joseph E and others},
  booktitle={17th USENIX Symposium on Operating Systems Design and Implementation (OSDI 23)},
  pages={663--679},
  year={2023}
}

@inproceedings{kwon2023efficient,
  title={Efficient memory management for large language model serving with pagedattention},
  author={Kwon, Woosuk and Li, Zhuohan and Zhuang, Siyuan and Sheng, Ying and Zheng, Lianmin and Yu, Cody Hao and Gonzalez, Joseph and Zhang, Hao and Stoica, Ion},
  booktitle={Proceedings of the 29th symposium on operating systems principles},
  pages={611--626},
  year={2023}
}

@article{zheng2023efficiently,
  title={Efficiently Programming Large Language Models using SGLang.},
  author={Zheng, Lianmin and Yin, Liangsheng and Xie, Zhiqiang and Huang, Jeff and Sun, Chuyue and Yu, Cody\_Hao and Cao, Shiyi and Kozyrakis, Christos and Stoica, Ion and Gonzalez, Joseph E and others},
  year={2023},
  publisher={arXiv}
}

@article{zheng2024sglang,
  title={Sglang: Efficient execution of structured language model programs},
  author={Zheng, Lianmin and Yin, Liangsheng and Xie, Zhiqiang and Sun, Chuyue and Huang, Jeff and Yu, Cody H and Cao, Shiyi and Kozyrakis, Christos and Stoica, Ion and Gonzalez, Joseph E and others},
  journal={Advances in neural information processing systems},
  volume={37},
  pages={62557--62583},
  year={2024}
}

@inproceedings{gao2024cost,
  title={$\{$Cost-Efficient$\}$ large language model serving for multi-turn conversations with $\{$CachedAttention$\}$},
  author={Gao, Bin and He, Zhuomin and Sharma, Puru and Kang, Qingxuan and Jevdjic, Djordje and Deng, Junbo and Yang, Xingkun and Yu, Zhou and Zuo, Pengfei},
  booktitle={2024 USENIX annual technical conference (USENIX ATC 24)},
  pages={111--126},
  year={2024}
}

@inproceedings{agrawal2024taming,
  title={Taming $\{$Throughput-Latency$\}$ tradeoff in $\{$LLM$\}$ inference with $\{$Sarathi-Serve$\}$},
  author={Agrawal, Amey and Kedia, Nitin and Panwar, Ashish and Mohan, Jayashree and Kwatra, Nipun and Gulavani, Bhargav and Tumanov, Alexey and Ramjee, Ramachandran},
  booktitle={18th USENIX symposium on operating systems design and implementation (OSDI 24)},
  pages={117--134},
  year={2024}
}

@inproceedings{zhong2024distserve,
  title={$\{$DistServe$\}$: Disaggregating prefill and decoding for goodput-optimized large language model serving},
  author={Zhong, Yinmin and Liu, Shengyu and Chen, Junda and Hu, Jianbo and Zhu, Yibo and Liu, Xuanzhe and Jin, Xin and Zhang, Hao},
  booktitle={18th USENIX Symposium on Operating Systems Design and Implementation (OSDI 24)},
  pages={193--210},
  year={2024}
}

@inproceedings{patel2024splitwise,
  title={Splitwise: Efficient generative llm inference using phase splitting},
  author={Patel, Pratyush and Choukse, Esha and Zhang, Chaojie and Shah, Aashaka and Goiri, {\'I}{\~n}igo and Maleki, Saeed and Bianchini, Ricardo},
  booktitle={2024 ACM/IEEE 51st Annual International Symposium on Computer Architecture (ISCA)},
  pages={118--132},
  year={2024},
  organization={IEEE}
}

@article{li2025continuum,
  title={Continuum: Efficient and robust multi-turn llm agent scheduling with kv cache time-to-live},
  author={Li, Hanchen and He, Runyuan and Mang, Qiuyang and Zhang, Qizheng and Mao, Huanzhi and Chen, Xiaokun and Zhou, Hangrui and Cheung, Alvin and Gonzalez, Joseph and Stoica, Ion},
  journal={arXiv preprint arXiv:2511.02230},
  year={2025}
}

@article{kipf2016semi,
  title={Semi-supervised classification with graph convolutional networks},
  author={Kipf, Thomas N and Welling, Max},
  journal={arXiv preprint arXiv:1609.02907},
  year={2016}
}

@article{velivckovic2017graph,
  title={Graph attention networks},
  author={Veli{\v{c}}kovi{\'c}, Petar and Cucurull, Guillem and Casanova, Arantxa and Romero, Adriana and Lio, Pietro and Bengio, Yoshua},
  journal={arXiv preprint arXiv:1710.10903},
  year={2017}
}

\end{document}